\documentclass[%
 reprint, nofootinbib,
 amsmath,amssymb,
 aps,
]{revtex4-2}

\usepackage{graphicx}
\usepackage{dcolumn}
\usepackage{bm}

\usepackage{hyperref}
\hypersetup{
    colorlinks=true,
    linkcolor=blue,
    filecolor=blue,      
    urlcolor=blue,
    citecolor = blue,
    pdfpagemode=FullScreen,
    }
\usepackage{natbib} 
\usepackage{xcolor}
\usepackage{float}
\usepackage{orcidlink}
\setcitestyle{authoryear,open={(},close={)}}

\begin{document}


\preprint{APS/123-QED}
\title{Systematic Uncertainties in the Measurement of Neutron lifetime \\ Using Lunar Prospector Neutron Spectrometer}

\author{Akshatha K Vydula\, \orcidlink{0000-0002-6611-2668}}
 \affiliation{School of Earth and Space Exploration, Arizona State University, Tempe, AZ 85281, USA}
 \affiliation{Space Science and Applications Group, Los Alamos National Laboratory, Los Alamos, NM, USA}
 \email[Correspondence email: ] {vydula@asu.edu}
 
\author{Daniel Coupland}
 \affiliation{Space Science and Applications Group, Los Alamos National Laboratory, Los Alamos, NM, USA}
 
\author{Katherine Mesick}
 \affiliation{Space Science and Applications Group, Los Alamos National Laboratory, Los Alamos, NM, USA}
 
\author{Craig Hardgrove}
 \affiliation{School of Earth and Space Exploration, Arizona State University, Tempe, AZ 85281, USA}

\date{\today}

\begin{abstract}
The lifetime of free neutrons measured in the lab has a long standing disparity of $\sim$9~s. A space-based technique has recently been proposed to independently measure the neutron lifetime using interactions between the galactic cosmic rays and a low atmosphere planetary body. This technique has not produced competitive results yet due to constraints of non-optimized data that contain large systematic errors. We use data from the neutron spectrometer on-board NASA's Lunar Prospector, and study two large systematics in the measurement of neutron lifetime: the lunar sub-surface temperature and the lunar surface composition. We use the HeCd and HeSn neutron spectrometer data when the spacecraft was in a highly elliptical orbit during the orbit insertion period. We report the neutron lifetime using four different models that each have different choices of surface temperature and composition. 5$^{\circ}$ \cite{prettyman2006elemental} and 2$^{\circ}$ re-binned \cite{wilson2021measurement} maps result in 777.6$\pm$11.7~s and 739.6$\pm$10.8~s respectively. For the 20$^{\circ}$ map, constant equatorial and a latitude-dependent temperature model result in 738.6$\pm$10.8~s and 767.3$\pm$11.2~s respectively. Increasing the complexities of the models accounting for the systematic effects increase the measured lifetime. However, the reported measurements are not competitive with the laboratory results due to large unaccounted systematics resulting from non-optimized measurements and modeling assumptions. This work serves as a study of systematic uncertainties for future neutron lifetime measurements using the space-based technique. We estimate the effect on the lifetime from the choice of
temperature model to be to be 28.7 $\pm$ 15.5~s, and choice of compositional map (for 20$^\circ$ and 5$^\circ$ maps) to be 10.3 $\pm$ 12.2~s.

\end{abstract}


\maketitle

\section{Introduction} \label{sec:intro}

The measurement of the lifetime of neutrons has been a project in progress for several decades, now actively being researched by many independent groups around the globe. Neutron decay is the simplest example of beta decay. A sufficiently precise neutron lifetime measurement thus offers an exquisitely clean probe of the underlying weak interactions, which places constraints on physics beyond the Standard Model \citep{wietfeldt2011colloquium}. A precise measurement of the mean lifetime of the neutron is also significant in understanding the rate of nuclear energy generation via the p-p chain reactions in stellar cores \citep{feldman1990technique} and are important in estimation of primordial abundance of helium in the stellar interiors. 

There are primarily two different methods of measuring neutron lifetime: beam and bottle experiments. The beam experiment measures the appearance of protons from the decay of a beam of neutrons \citep{wietfeldt2011colloquium}. \cite{robson1951} made one of the first measurements of the neutron lifetime using the beam method, with the most precise recent measurement being $\tau_n = 887.7 \pm 1.2_{stat} \pm 1.9_{syst}~s$ by \cite{yue2013improved}. Further, a more precise value is expected with $<$1~s of uncertainty in the next few years \citep{hoogerheide2019progress}. The bottle method measures the disappearance of neutrons in a confined system as a function of time. \cite{kosvintsev1980use} gave the proof of principle in 1980 and reported $\tau_n = 875 \pm 95~s$. The uncertainties in the measurement were significantly improved within a decade to $\tau_n = 887.6 \pm 3~s$ as reported by \cite{mampe1989neutron} and has been getting better with the most recent measurement being $\tau_n = 877.75 \pm 0.28_{\text{stat}} +0.22 / {-0.16_{\text{syst}}}~\textsl{s}$ by \cite{gonzalez2021improved}. The  most precise beam and bottle methods differ by over $4\sigma$ \cite{rajan2020meta}. This discrepancy is most likely attributed to potentially unresolved systematic uncertainties in one or both measurement techniques.

A third method of measuring neutron lifetime is now actively being explored. A space-based measurement technique makes use of the free neutrons that are naturally produced by interactions between Galactic Cosmic Rays (GCRs) and a planetary body, such as the Moon. The lowest energy ($<$ 0.5 eV) of these neutrons have a velocity less than 10 km/s, and can thus undergo beta decay in the distances reasonably sampled by an orbiting satellite. The neutron lifetime can then be measured based on the disappearance of these neutrons with distance from the planetary body. These same neutrons have been measured by many planetary science missions because they provide insight into the surface composition of planetary bodies. Thus while no dedicated space-based neutron lifetime measurement has been performed, initial studies are possible with existing planetary science data. A feasibility study of this method was first done by \cite{wilson2020space}, estimating the $\tau_n = 780 \pm 70~s$ using flybys of Venus and Mercury by NASA's MESSENGER spacecraft. A more recent measurement of $\tau_n = 887 \pm 14~s$ has been reported by \cite{wilson2021measurement} using data from the Lunar Prospector (LP) mission. It is pointed out in \cite{wilson2021measurement}  that the biggest sources of systematic uncertainties in such measurements are expected to be the choice of models describing the lunar composition and latitudinal temperature variations that dictate the thermal neutron flux.



The LP mission was a planetary science mission and was not designed for a neutron lifetime measurement; therefore, we do not expect the uncertainty reported here to compare with the accuracy and precision of current beam or bottle measurement techniques. This analysis serves as a study of systematic uncertainties for the space-based technique for future neutron lifetime measurements. While several sources of systematic uncertainties exist in the space-based measurement technique, in this paper, we consider uncertainties in the neutron lifetime measurement due to the choice of a surface composition model and a temperature model. We test four different models that capture these systematics and report a neutron lifetime for each case. The energy spectrum of the neutrons that leak from a planetary surface is dependent on the local surface composition \citep{mesick2018benchmarking}, which is why neutron measurements are an effective tool in planetary science. 

Specifically, we study the effects of temperature of the lunar subsurface and the resolution of composition maps on the computation of the neutron lifetime. We consider surface compositions derived from orbital measurements at three spatial resolutions. At the smallest spatial resolution (2$^{\circ}$ in latitude and longitude) we  limit ourselves to considering five distinct compositions, following the work of \cite{wilson2021measurement}. At coarser resolutions (5$^{\circ}$ and 20$^{\circ}$ latitude with roughly equal spatial size in longitude) we use compositions derived from LP Gamma-ray Spectrometer (GRS) measurements, which are sampled at similar spatial and depth scales as the LP Neutron Spectrometer (LPNS). As the thermal neutron emission may be impacted by the temperature of the lunar subsurface, we consider both constant temperature and a latitude-dependent temperature model specifically developed for the analysis of LPNS data. Starting from these lunar surface definitions, we construct a simulation chain that models LPNS data as a function of neutron lifetime during the LP early-mission high-altitude orbits. We compare the simulated neutron current to the measured LPNS data to arrive at the neutron lifetime. We compare the values derived from different surface treatments and discuss the uncertainty in the measurement.


The rest of the paper is organized as follows. We describe the LP data and simulations in section \ref{sec:data_sim}. The on-board instrument used is described in section \ref{sec:ns}, choice of models to study the effects of lunar surface compositions and temperatures in section \ref{sec:comp}, \texttt{MCNP} simulations in section \ref{sec:mcnp} and the neutron propagation model in section \ref{sec:prop_mod}. We discuss the results of effects of lunar composition in section \ref{sec:comp_effects} and those of sub-surface temperatures in section \ref{sec:temp_effects}. We then explain the statistical approach used in computing the results using our models and the LP data in section \ref{sec:stats}. We layout future work in section \ref{future} and give our conclusions in section \ref{sec:conclusion}.

\section{Data and Simulations}
\label{sec:data_sim}
The LP mission was in a low altitude, near-circular lunar orbit from January 1998 to July 1999 \citep{feldman1999lunar}. It was designed to make measurements to map elemental abundances to depths of about 20~cm of the lunar terrain. On-board were five science instruments to make various lunar measurements - the Gamma-ray spectrometer (GRS), the Neutron spectrometer (NS), the Magnetometer/electron reflectometer, the Alpha particle spectrometer (APS) and the Doppler gravity experiment (DGE) \citep{binder1998lunar}. A total of 11 major and trace elements were mapped using the NS and GRS during its approximately year and a half science mission. While we use the results of this full analysis to inform the simulated surface, for the neutron lifetime analysis we use only the NS data from LP's orbit insertion period of about 3 days. During this 3 day flight period, the spacecraft went through multiple stages of orbit corrections starting from a very elliptical orbit to a circular one. The initial orbit insertion data is appropriately suited for the measurement of neutron lifetime, as we record the disappearance of thermal neutrons as a function of altitude (ranging from 85~km to $\sim$16900~km). For the lowest-energy neutrons measured by the NS, the time it takes to reach these altitudes is comparable to the neutron lifetime. The measurements used in this analysis include the count rates from the two detector elements of the NS and the satellite ephemeris, which includes position and orientation. The data reduction techniques are described in detail by \cite{binder1998lunar}. The full data reduction described by \cite{maurice2004reduction} was not performed on the orbital insertion data and is not directly extendable to these data due to empirical corrections that are specific to the orbits and orientations of the circular orbit phases of the mission.

\subsection{Neutron Detector}
\label{sec:ns}
LP consists of two geometrically identical neutron detector proportional counter tubes, each of 20~cm in length and 5.7~cm in diameter \citep{feldman2004gamma}. The tubes are filled with $^3$He gas at 10~atm pressure and covered with a 0.63~mm thick sheet of either Cd or Sn. The Cd absorbs low energy neutrons, so the ``HeCd'' detector is sensitive to neutrons of energy $>0.4$~eV (epithermal neutrons) while the ``HeSn'' detector is sensitive to neutrons of all energies (thermal and epithermal neutrons). The data product from these detectors is a 32-bin pulse amplitude histogram that is produced at a 32 second cadence. We sum the bins in the second half of the pulse height spectra, bins corresponding to higher energy deposition events in the detector to reduce the contribution of non-neutron background interactions, corresponding to energy bins 17 to 32. This produces the neutron counts per 32 seconds data product for the HeCd and HeSn detectors and the corresponding uncertainties in measurements. This choice was made to avoid the background noise and to achieve higher signal-to-noise in the pulse height spectrum \citep{maurice2004reduction}. The background rates in this data were assessed as the average measured rate far from the Moon in the days before lunar orbital insertion. These background rates are 134.9 and 127.8 counts/32-s for the HeCd and HeSn detectors, respectively. Backgrounds are the result of GCR spallation on the spacecraft itself, and so they decrease at lower altitudes as the Moon blocks incident GCRs. The decrease in background is linear with the solid angle subtended by the Moon as a function of altitude, which is 35\% at the lowest altitudes in this dataset. Background-subtracted rates are shown in the top panel of Figure~\ref{fig:data_vs_model}. To get the thermal neutron counts (energy range of $<$0.4~eV), we subtract the HeCd count rates from the HeSn count rates.

\subsection{Simulating Surface Composition \& Temperature}
\label{sec:comp}
We construct simulated neutron counts in three steps: first by simulating the neutron emission from a given surface composition and temperature, then by mapping these compositions onto the lunar surface and calculating their contribution to the neutron current at a given point in the LP orbit, and finally by evaluating the response of the two NS detectors to the current to produce a count rate that is directly comparable to the detector count rates.

\cite{wilson2021measurement} derived five lunar compositions from \cite{peplowski2016geochemistry} and \cite{jolliff2000major}, which are re-binned into a 2$^{\circ}$ resolution map for this work. These compostions are the procellarum KREEP terrane (PKT), South Pole Aitken terrane (SPA), feldspathic highland terrane (FHT), pure anorthosite terrane (PAN) and non-PKT maria (nPKT). For elemental mass fraction of these composition, refer to Table II in \cite{wilson2021measurement}. Although this map has been used to estimate the value of neutron lifetime, it is a simplified map, and fails to capture the complexities of the lunar terrain. Here, in addition to the five listed compositions, we also model the neutron current from the lunar surface at two more resolutions derived from \cite{prettyman2006elemental} - (i) a 5$^\circ$ resolution map showing 1790 different compositions and (ii) a 20$^\circ$ resolution map showing 114 different compositions. The 5$^{\circ}$ binned composition map comes from the NASA Planetary Data System (PDS) \citep{LPPDS} and contains weight percent (wt\%) or ppm of Mg, O, Al, Si, Ca, Ti, Fe, K, Th, and U. Additional higher resolution (2$^{\circ}$) binned maps for H and Sm are included from the PDS. To form a complete set, the H and Sm were resampled to 5$^{\circ}$ and 20$^\circ$ resolutions. Further, Gd, an important neutron absorbing element, was included by scaling the amount of Sm by a constant factor of 1.3, adopted from \cite{hidaka2000neutron}. We note that, although both the 5$^{\circ}$ and 20$^{\circ}$ maps are obtained from the PDS, the 20$^{\circ}$ map is not directly derivable from the 5$^{\circ}$ map. Since the post-processing involves various steps including spatial averaging, interpolation, noise reduction among others, we use the available published maps. 


To study the effects of latitude-dependent surface temperature, we performed simulations for two different cases using the surface composition maps at 20$^{\circ}$ resolution reported in \cite{prettyman2006elemental}: (i) constant equatorial temperature of 240~K and (ii) latitude-dependent subsurface lunar temperature reported by \cite{little2003latitude}. 
The temperature (in $K$) is given by: 

\begin{equation}
    \label{temperature}
    T = \begin{cases}
    
    100  & \text{if } |lat| \geq 88.5^o \\
    250 cos^{1.25}(lat) & \text{if }  |lat| < 88.5^o 
  \end{cases}
\end{equation}
The temperature is modified using the \texttt{TMP} card in \texttt{MCNP}, which accounts for temperature effects on the elastic interaction cross sections assuming a free gas model. It also accounts for the thermal motion and modified collision kinematics reported in \cite{werner2018mcnp6}.

\subsection{Neutron Emission Simulations}
\label{sec:mcnp}
\texttt{MCNP}\textsuperscript{\textregistered} is  Los Alamos National Laboratory's general-purpose Monte Carlo N-Particle code\footnote{\href{https://mcnp.lanl.gov/}{\texttt{https://mcnp.lanl.gov/}}} that can be used for detector design analysis and particle transport analysis \citep{werner2018mcnp6}. We collect neutron current tallies on a spherical shell above the lunar surface. This, combined with the NS detector response gives the neutron count rates. For this work, \texttt{MCNP} version 6.2 was used.

To simulate the GCRs we use the particle flux equations in \cite{usoskin2017heliospheric} and the reported mean solar modulation during the orbital insertion of LP (Jan 1998) of 505~MV, which results in a GCR proton flux of $J_p$=3.575~particles/cm$^2$-s and $\alpha$ particle flux of $J_{alpha}$=0.323~particles/cm$^2$-s. We simulate only protons, but account for the contribution of $\alpha$ flux using the scaling from \cite{mckinney2006mcnpx} of a factor of $R=3.8$ more neutrons produced per incident $\alpha$ than per incident proton.  Thus the total equivalent omni-directional proton flux $J_{eq,p} = J_p +RJ_{alpha}$ is 4.8~p/cm$^2$-s. Since our \texttt{MCNP} simulations start with a $2\pi$ surface instead of total omni-directional flux, the incident flux is a factor of 4 lower \citep{mckinney2006mcnpx,mesick2018benchmarking}, resulting in an incident simulated proton flux of $J_{p,inc}$= 1.2~particles/cm$^2$-s.

We tally the neutrons emitted after the interaction of GCRs with the lunar surface using \texttt{MCNP} current tallys and normalize for source flux ($J_{p,inc}$) and tally area. We tested the angular distributions of a few widely-varying compositions and found that the angular distributions did not vary with composition. Most MCNP simulations were then performed without tallying the emission as a function of angle, which improved the convergence of the simulations with lower computational requirements. For each of the compositions tested, the normalized neutron current integrated over the thermal peak ($<$0.4~eV, when neutrons have slowed to an energy/velocity equal to a temperature dependent thermal equilibrium) is noted to be $\approx$0.2~particles/cm$^2$-s.

\begin{figure}[h]
	\includegraphics[width=\columnwidth]{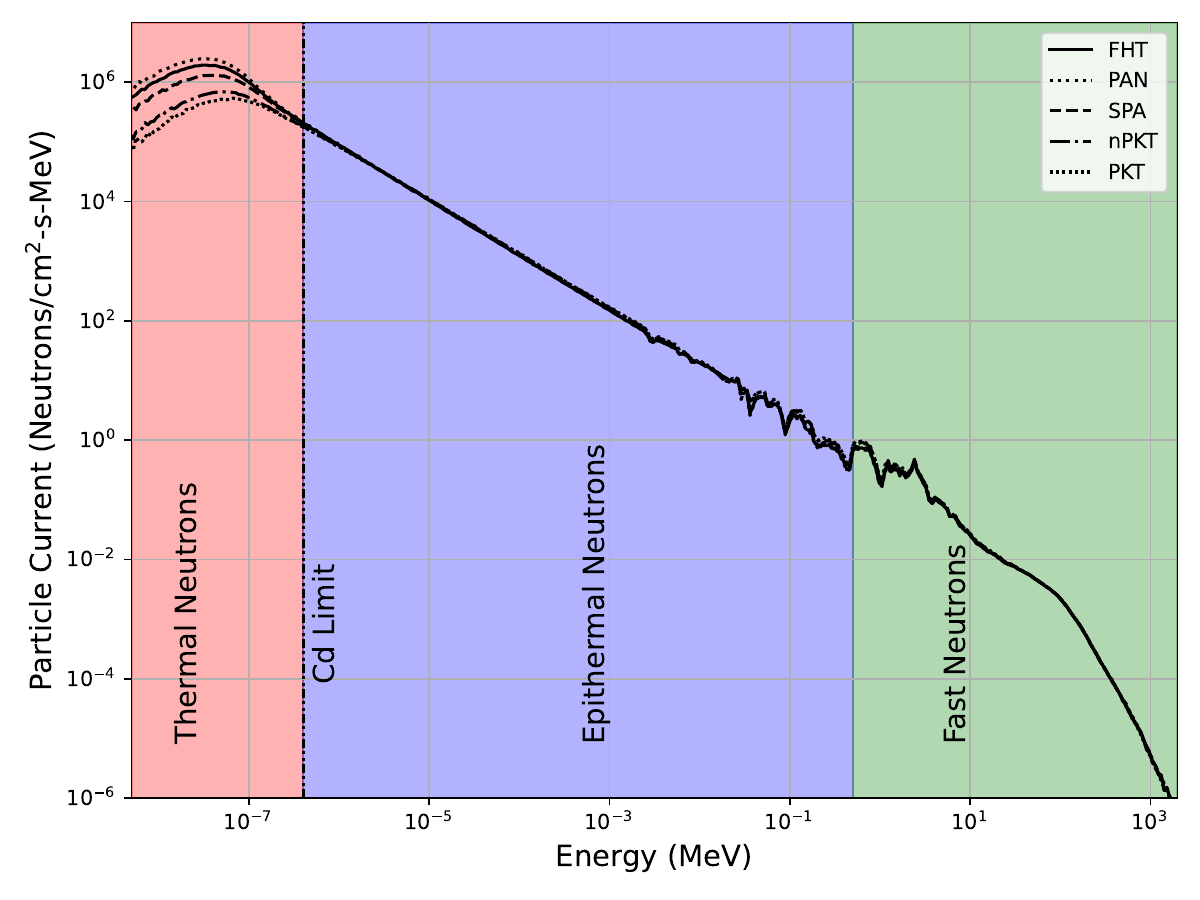}
    \caption{Comparison of normalized current tallys from PKT, nPKT, PAN, SPA and FHT compositions on a spherical surface at an altitude of 10~km above the lunar surface. The plot also shows the Cd limit of 0.4~eV and the resulting distinction between thermal, epithermal and fast neutrons. }
    \label{fig:normalized_mcnp}
\end{figure}

Figure \ref{fig:normalized_mcnp} shows the simulated neutron current for the five compositions reported in \cite{wilson2021measurement} at an altitude of 10~km above the lunar surface. The neutrons initially generated by the GCR interactions have energies on the order of 10s-100s of MeV \citep{mesick2018benchmarking}. The neutrons lose energy through elastic and inelastic scattering, eventually ending up in thermal equilibrium with the surface. The spectrum of neutrons emitted from the surface depends on the surface composition through the likely energy loss per scatter and the likelihood of absorption. The population of neutrons within the thermal, epithermal and fast regions of the neutron energy spectrum are counted to give insights into the surface composition for planetary science measurements. We observe a 20\% variation in the epithermal and fast neutron current; however, these differences are not easily discernible in Figure~\ref{fig:normalized_mcnp} due to the large range on the y-axis.


\texttt{MCNP} is designed to compute neutron current by creating surfaces through which neutrons cross after they are released from a planetary surface, such as the Moon. This approach assumes that the neutron escapes the gravitational field of the planetary body. However, low-energy neutrons are gravitationally-bound and can return, interact with the surface, and be re-emitted, which increases the neutron current at low energies\citep{feldman1989gravitational}. It may therefore be necessary to account for the low-energy neutrons with an energy less than 0.0295~eV, that do not escape lunar gravity. These effects can be accounted for using the \texttt{FIELD} card in \texttt{MCNP}6.2+ and \texttt{MCNPX}, which were implemented based on particle physics discussed in \cite{feldman1989gravitational}. However, we tested the effects of gravitationally bound downward traveling neutrons using the 5$^\circ$ map and found that \texttt{MCNP} 6.2 does not decay neutrons, resulting in an over-prediction of the gravity effects. We estimate the effect of ignoring gravitational bound neutrons to be less than 0.5\% on the total thermal neutron flux for lunar compositions with the low water content (tens of ppm) used in this study. We therefore disabled the \texttt{FIELD} card in our simulations.


It is established that the neutron spectrum produced using the compositions as measured by GRS does not match with the neutron spectrometer data \citep{peplowski2013compositional}. To account for this, we scale the elemental mass fraction of oxygen such that the weighted sum of the elemental microscopic neutron absorption cross sections matches the measured macroscopic neutron absorption cross section. The data for macroscopic absorption cross-section from the science-phase of the LP mission is obtained from \cite{feldman2000chemical} and GRS measurements are obtained from the PDS \citep{LPPDS}. The microscopic neutron absorption cross sections for each of the abundance-weighted sum of the elemental constituents are obtained from \cite{prettyman2006elemental}. We use the correction technique described in Equation 2 of \cite{peplowski2013compositional}:

\begin{equation}
    \Sigma_a = N_A \sum_i \Big(\frac{w_i}{A_i} \Big) \sigma_i~,
\end{equation}
where $N_A$ is Avogadro's number, $w_i$ is the elemental weight fraction, and $A_i$ is the atomic weight. This correction factor for some compositions was higher than others, but less than 30\% overall. The corrected compositions, after scaling elemental mass fraction of oxygen were used as inputs to the \texttt{MCNP} simulations.

\begin{figure*}[t!]
	\includegraphics[width=\textwidth]{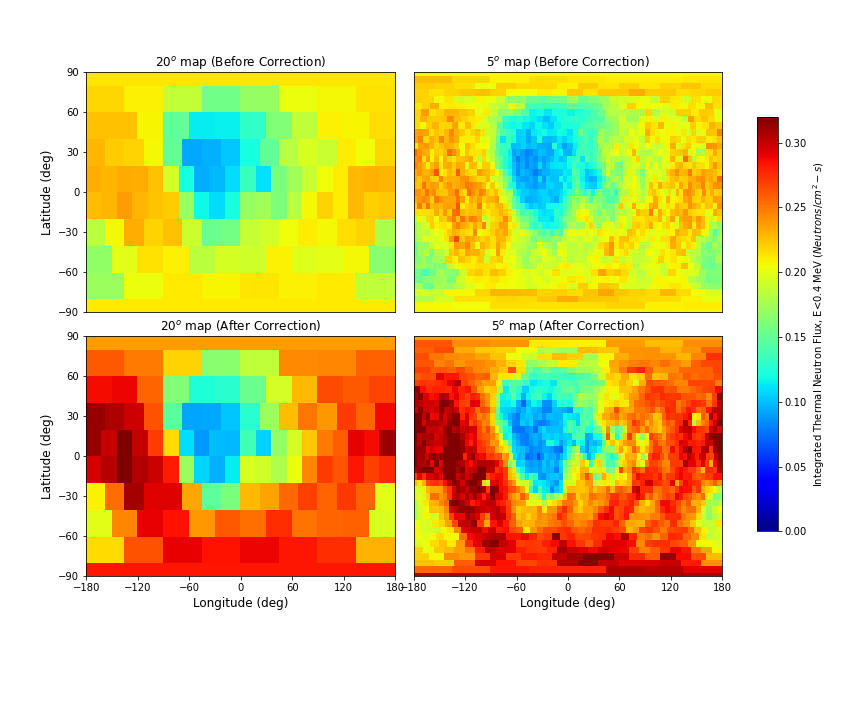}
    \caption{Integrated thermal neutron current for composition maps from \cite{prettyman2006elemental}, before and after composition correction. Correction was applied based on Equation 2 of \cite{peplowski2016geochemistry} using microscopic and macroscopic neutron absorption cross sections.  \textit{(Left:)} 20$^{\circ}$ maps and \textit{(Right:)} 5$^{\circ}$ maps.}
    \label{fig:integrated_flux}
\end{figure*}

Figure \ref{fig:integrated_flux} shows the integrated thermal neutron current for the 20$^{\circ}$ and 5$^{\circ}$ resolution maps before and after correction. An overall increase in integrated thermal neutrons is noted.

\subsection{Propagation Model}
\label{sec:prop_mod}



\label{comp_prop_model}
We start with neutrons from the spallation on the lunar surface, and use particle kinematics that take into account the ballistic trajectories due to gravity. The influence of lunar gravity is significant for low-energy neutrons used in the measurement of neutron lifetime. Neutrons with energies below 0.0295~eV that exit the lunar surface will become gravitationally bound. If these neutrons do not undergo beta decay during flight, they will return to the surface. On the contrary, higher-energy neutrons will follow hyperbolic trajectories, trading kinetic energy for potential energy as a function of radial distance. The orbits of these neutrons originate at all points on the lunar surface and intersect with a given satellite location. The closed-form solutions for neutron flux at a given satellite position due to neutron propagation in a spherically-symmetric gravitational field were presented by \cite{feldman1989gravitational} (Feldman formalism). The solutions use the incident energy, angle, flux density, and transit time as a function of emission energy, angle, and radius. These are briefly described in Appendix \ref{appendic:particle_kinematics}, including the correction of a sign error as noted in \cite{mesick2020new}. This approach, commonly used in planetary science neutron analyses, does not account for the location of neutron emission. It assumes that the neutrons reaching a given satellite location originate from a single surface composition. This assumption is valid for planetary science analyses focused on probing surface composition and reflects the spatial resolution of those studies.

However, in this analysis, the satellite's altitude is high enough to sample multiple composition units on the surface, thus supplementing the Feldman formalism is essential, such that the neutron flux can be mapped to its emission location. Specifically, we apply the Feldman formalism to generate a grid of neutron energies, angles, and transit times originating from the surface, along with their corresponding values at a given satellite location, for both upward-traveling neutrons and gravitationally-bound neutrons returning to the surface. This provides an analytical treatment of the changing neutron flux density from the surface-to-point problem. We then describe the neutron trajectories on that grid as orbital motion (following e.g. \cite{Curtis_OMES2005}), calculating the orbit eccentricity and the true anomaly at both the satellite and the surface. The true anomaly for neutrons with inward trajectories intersecting the spacecraft is described in Appendix \ref{app:true_anomaly}.  Since the true anomaly difference is simply an angular difference, this allows us to trace the neutron orbit from the satellite location back to its intersection with the surface. 

Since the satellite's velocity is comparable to that of the neutrons, it must be subtracted from the neutron velocity to obtain the neutron velocity in the rest frame of the detectors. We note that the spacecraft velocity provided by the mission ephemeris is in a global cartesian coordinate system with the z-axis aligned with the rotation axis of the Moon. To account for this, the neutron velocities are transformed from the coordinate system of the Feldman formalism, in which all angles are relative to local zenith, to the global coordinate system. This is described in \ref{appendix:transforms}. After subtracting the satellite velocity, we perform a final coordinate transformation to the spacecraft's reference frame. The spacecraft spins at approximately 12 rotations per minute, so only the spin axis is relevant. During most of the main science phase of the mission, the spin axis was roughly aligned (or anti-aligned) with the Moon's rotation axis. However, this alignment was not in place during the orbital insertion period used in this analysis.  We use the spacecraft attitude information from the ephemeris, which specifies the direction of the spacecraft spin axis as a right ascension (196-197$^\circ$) and declination (58-59$^\circ$) in the ECLIPJ2000 reference frame. We retrieve the coordinate axes of the Moon in the same reference frame using the \texttt{SpiceyPy} \citep{annex2020spiceypy} wrapper of the \texttt{SPICE} toolkit \citep{acton1996ancillary,acton2018look}, and transform the neutron velocity into the reference frame of the LP spacecraft.


\begin{figure}[H]
	\includegraphics[width=\columnwidth]{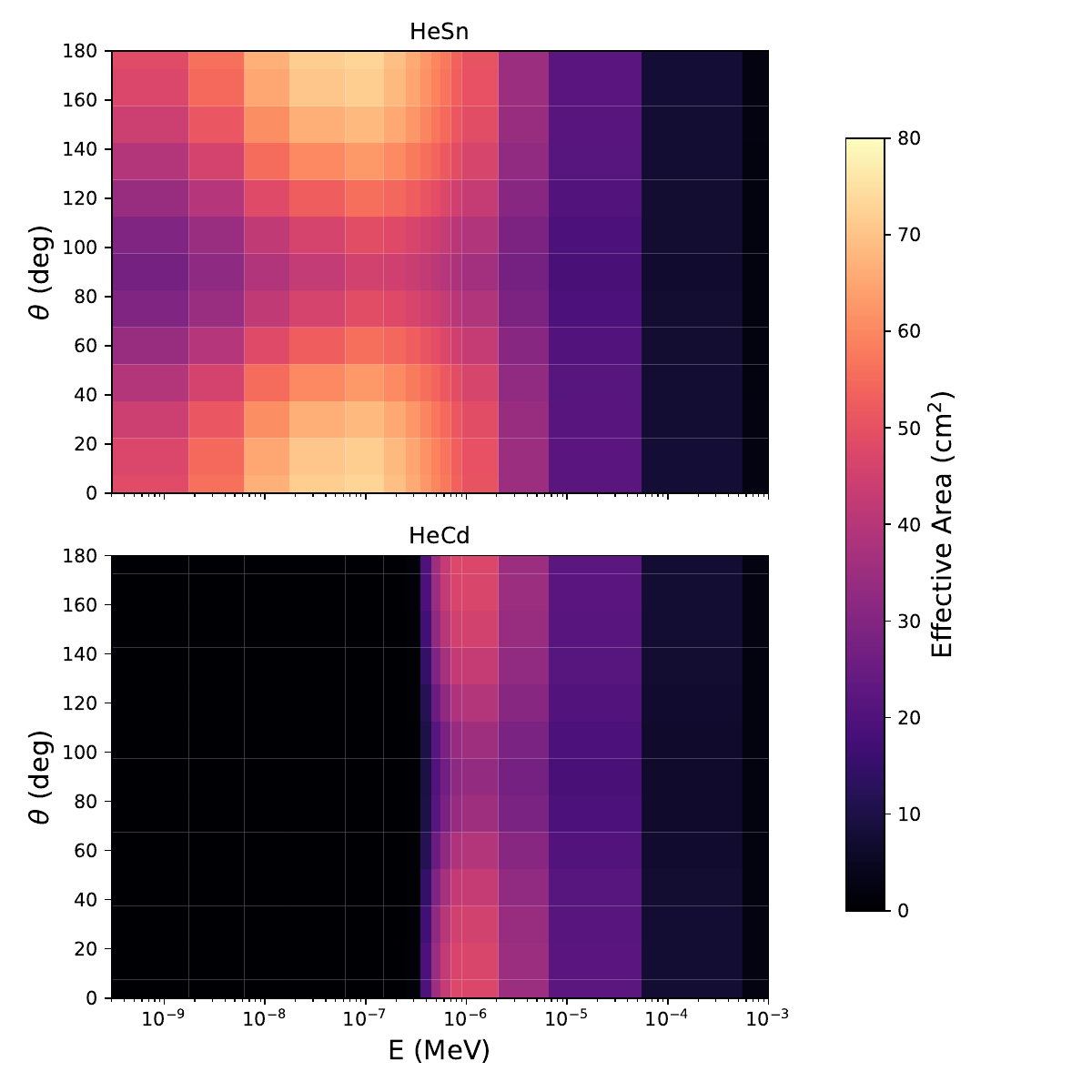}
    \caption{Effective areas of HeCd and HeSn detectors assuming the length of detector to be 16.72~cm and using a 0.54~MeV energy deposition threshold. Simulations were performed using \texttt{Geant4} version 10.07.}
    \label{fig:effective_area}
\end{figure}

The detector response was simulated using \texttt{Geant4} version 10.07 \citep{allison2016recent,allison2006geant4, agostinelli2003geant4}, with a detector geometry derived from the original LPNS instrument description \citep{feldman1999lunar}. More recent measurements \citep{peplowski2020position} have shown that similar cylindrical He$^3$-filled gas proportional counters have a position-dependent response, shortening the region that detects incident neutrons. With this correction, the best current estimate of the active region of the LPNS detectors is 16.72 cm long rather than the original 20 cm specification (David Lawrence, email message to co-author, July 6, 2022). A 0.54 MeV energy deposited threshold was used in the simulation to correspond to energy bins 17 to 32 in the measurement data. This choice ensures that the pulse height spectra in the 32-bin pulse amplitude histogram is captures and the contribution of non-neutron background interactions is minimal. Further, only the proportional counter tubes and their immediate cladding were included in the simulation geometry. 

Lunar Prospector was spin-stabilized, rotating around its axis with a period of roughly 5 seconds, which is much shorter than the 32 second integration time of the primary data set. The detector response was thus computed spin-averaged around the spacecraft rotation axis. The spacecraft bus itself shadows the detectors in some of the geometries. This effect was corrected empirically in the LPNS primary mission data \citep{maurice2004reduction} rather than being modeled. However, those results are not applicable to the orbital insertion data used in this study. Thus, this effect is not corrected in our analysis. The results of \cite{maurice2004reduction} suggest that the impact of this effect is between 0\% and $\approx$ 30\% shadowing depending on the orientation of the spacecraft relative to the lunar surface. The final spin-averaged detector responses are shown in Figure~\ref{fig:effective_area} for each detector, expressed as the detector effective area, i.e., the size of an equivalent perfectly efficient detector. Neutron flux was convolved with the detector response to obtain the detector count rates. The simulated detector count rate, for each of the two detectors, is thus the product of the incident neutron flux and effective area integrated over all energies and angles.

\begin{figure}[h!]
	\includegraphics[width=\columnwidth]{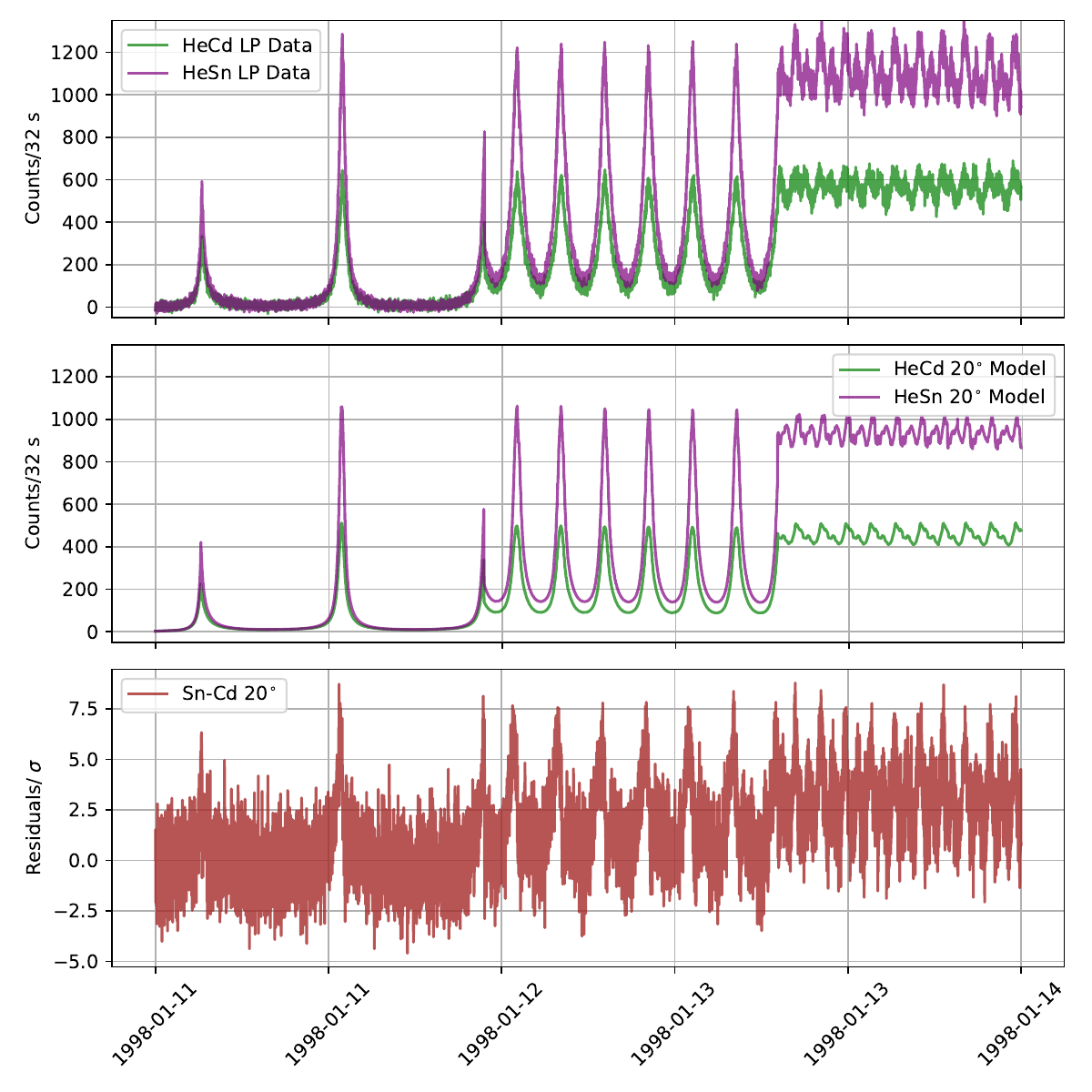}
    \caption{\textit{(Top):} HeCd and HeSn detector count rates from LPNS data and \textit{(middle):} those determined from the model using 20$^{\circ}$ resolution composition map. and \textit{(bottom):} Residuals between data and model.}
    \label{fig:data_vs_model}
\end{figure}

We subtract the HeCd count rates from the HeSn count rates to obtain the thermal neutrons (E$<$ 0.4~MeV), which are used as the model parameter to compute the neutron lifetime. The thermal neutron residual is computed through a simple subtraction of thermal neutrons using the measured and modeled detector count rates. Figure \ref{fig:data_vs_model} shows simulated and measured HeCd and HeSn detector count rates from LP data using surface composition of 20$^{\circ}$ resolution map, and the corresponding thermal neutron residuals between data and model. The computed detector count rates are assumed to be Poisson distributed, and the uncertainties ($\sigma$) are computed using 16 channel count rates containing the pulse height spectra of the 32 channel LP detector counts \citep{maurice2004reduction}.  We note that the residuals have low variance ($\sim$ 1.7 $\sigma$ for the 20$^\circ$ map), however retains structures when the spacecraft is in an active orbit maneuver. 

\section{Effects of Lunar conditions on Neutron Lifetime}
\label{effects}
Computing neutron lifetime needs precise modeling of the lunar terrain, including knowledge of the lunar surface composition and temperature. Here, we quantify the effects of surface composition by using different resolutions of the composition maps obtained from the LP data. We also quantify the effects of lunar sub-surface temperature. Both the effects are modeled using \texttt{MCNP} simulations, which are used in the neutron propagation pipeline described in Section \ref{sec:prop_mod}.

\subsection{Effects surface composition and resolution}
\label{sec:comp_effects}

\begin{figure}[h]
	\includegraphics[width=\columnwidth]{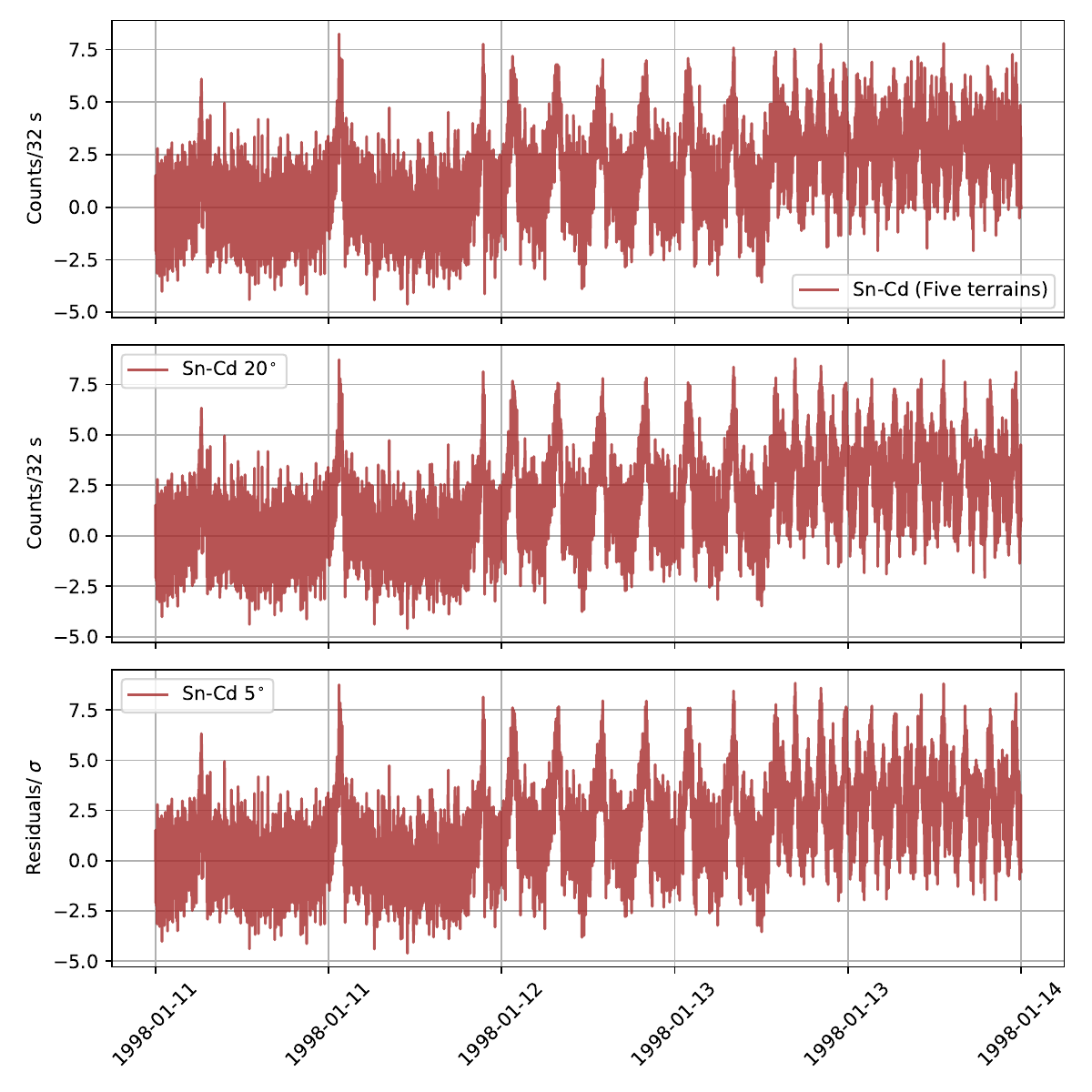}
    \caption{Thermal neutron residuals in LP data and model computed using lunar surface composition at three different resolutions, at a nominal value of $\tau_n$ = 880~s. \textit{(Top):} Five compositions used in \cite{wilson2021measurement}, \textit{(middle):} 114 different compositions at 20$^\circ$ resolution, and \textit{(bottom):} 1790 composition at 5$^\circ$ resolution, both obtained from \cite{prettyman2006elemental}.}
    \label{fig:composition_effects}
\end{figure}

To quantify the effects of the lunar sub-surface composition, the neutron current was modeled in \texttt{MCNP} using three different resolutions of composition maps, as described in Section \ref{sec:comp}. Figure \ref{fig:composition_effects} shows the residuals in thermal neutrons between LP data and each of the three composition maps. All three resolutions of the maps have low statistical RMS in thermal neutron residuals. The composition map adopted from \cite{wilson2021measurement} that uses five compositions re-binned at 2$^\circ$ resolution, produces the smallest RMS. These variances are reported in Table \ref{tab:composition_effects}. We note that, although a simpler composition map has the least RMS in residuals, both 5$^{\circ}$ and 20$^{\circ}$ maps capture the variations in lunar terrain better than the five compositions. We also note that the composition maps are measurements made by LP during its mission period in a circular orbit, which could imply biases in this analysis. During the orbit insertion period considered in this analysis, the closest spacecraft approach is $\sim$85~km, while the farthest approach is $\sim$16,900~km. Using 1.5 times the altitude as an estimate for the spatial footprint, we calculate the spatial footprint to be 127.5~km, or 4.2$^\circ$, at the closest approach. At this distance, the half-angle to the limb of the Moon is 17.6$^\circ$, suggesting a significant reduction in contributions due to the non-uniform neutron emission angles. However, at the farthest distance, the half-angle increases to 84.7$^\circ$, indicating that neutron emission is sourced from nearly half of the Moon’s surface. A more detailed and independent understanding of the lunar terrain and neutron emission is necessary to fully understand the effects of composition maps on the neutron lifetime.  

\begin{table}[]
    \caption{Residuals in thermal neutrons between LP data and models obtained using three different resolution maps. }
    \label{tab:composition_effects}
    \centering
    \begin{ruledtabular}
    \begin{tabular}{c c c}
        Map Resolution & No. of compositions & Thermal neutron\\
        & & Residuals ($\sigma$) \\
        \hline
        \\
        2$^\circ$  & 5 & 2.51\\
        20$^\circ$  & 114 & 2.68\\
        5$^\circ$  & 1790 & 2.64\
        
    \end{tabular}
    \end{ruledtabular}
\end{table}

\subsection{Effects of surface temperature}
\label{sec:temp_effects}

\begin{figure}[h]
    \centering
    \includegraphics[width=\columnwidth]{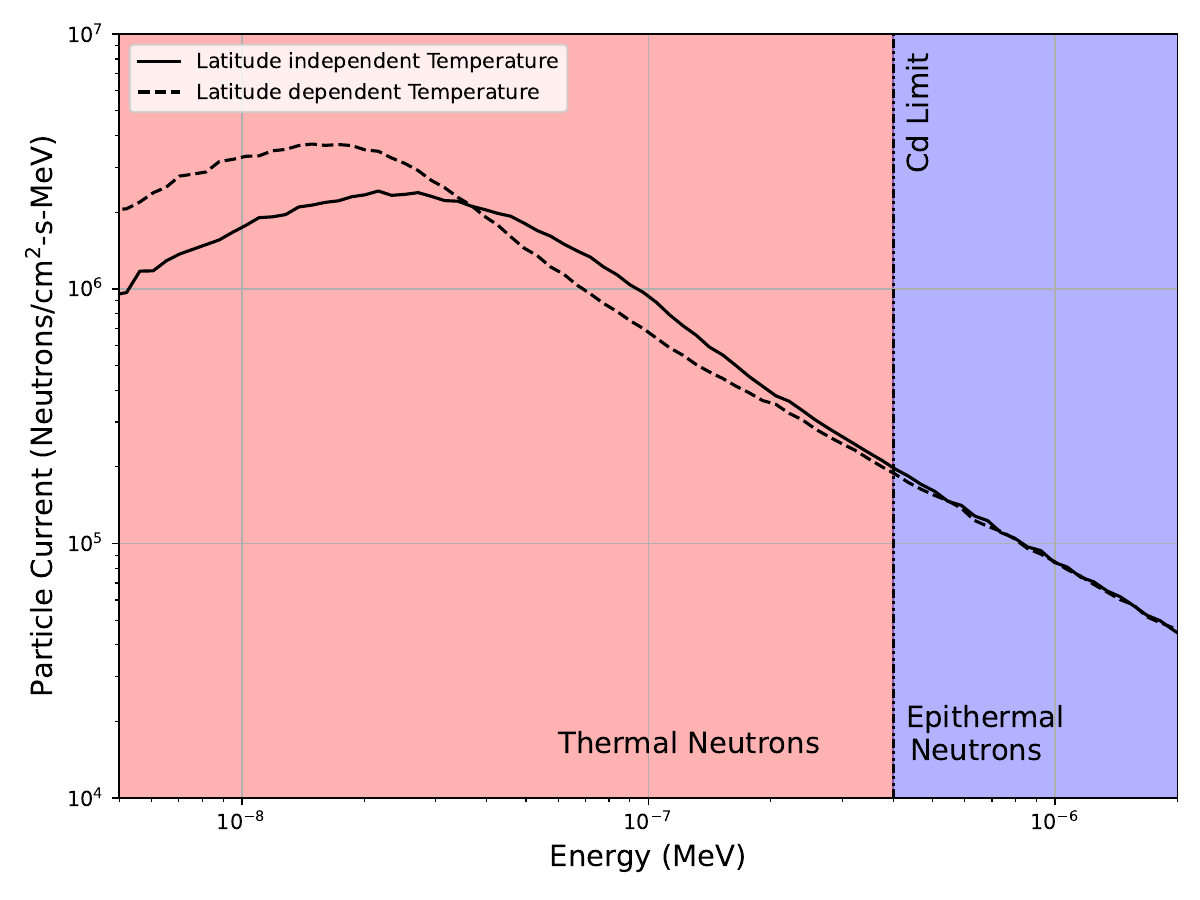}
    \caption{Neutron spectra obtained from \texttt{MCNP} simulations showing lunar subsurface temperature effects. For a model that accounts for the latitude dependent temperature variances, a higher neutron current is seen at a lower thermal resonance.}
    \label{fig:temperature_spectra}
\end{figure}

To quantify the effects of the sub-surface temperature on neutron lifetime, two different cases are used in \texttt{MCNP} modeling - a (i) constant equatorial temperature at all latitudes and (ii) a latitude dependent temperature where poles are cooler than the equator. Figure \ref{fig:temperature_spectra} shows the neutron spectra for the two cases, using one of the compositions picked from the 20$^{\circ}$ resolution composition map. For a constant equatorial temperature case, the poles are assumed to be hotter than they actually are. This produces a lower particle current and the thermal resonance is reached at a slightly higher energy. When the temperature is modeled as a function of the latitude, the thermal resonance is achieved at a relatively lower energy but with a relatively a higher particle current. Overall, higher number of thermal neutrons are seen when the effects of latitude dependent temperature is modeled.

We note that in the low energy regimes ($<0.2~eV$), the neutron current is relatively lower at the poles for the temperature varying case compared to the constant equatorial temperature due to lower subsurface temperatures. To quantify the effects of subsurface temperatures, we calculate the correlation coefficients between these two cases, given by:

\begin{equation}
    \label{correlation_coeff}
    \rho_{C,V} = \frac{\mathbb{E}[(f_C -\mu_{C})(f_V-\mu_{V})]}{\sigma_C \sigma_V}
\end{equation}
where $f_C$ and $f_V$ represent normalized neutron current for constant equatorial temperature and latitude dependent subsurface temperature \texttt{MCNP} outputs respectively, and $\sigma_C$ and $\sigma_V$ represent corresponding standard deviations; and $\mu_{C}$ and $\mu_{V}$ are mean of $f_C$ and $f_V$ respectively.

\begin{figure*}[]
	\includegraphics[width=\textwidth]{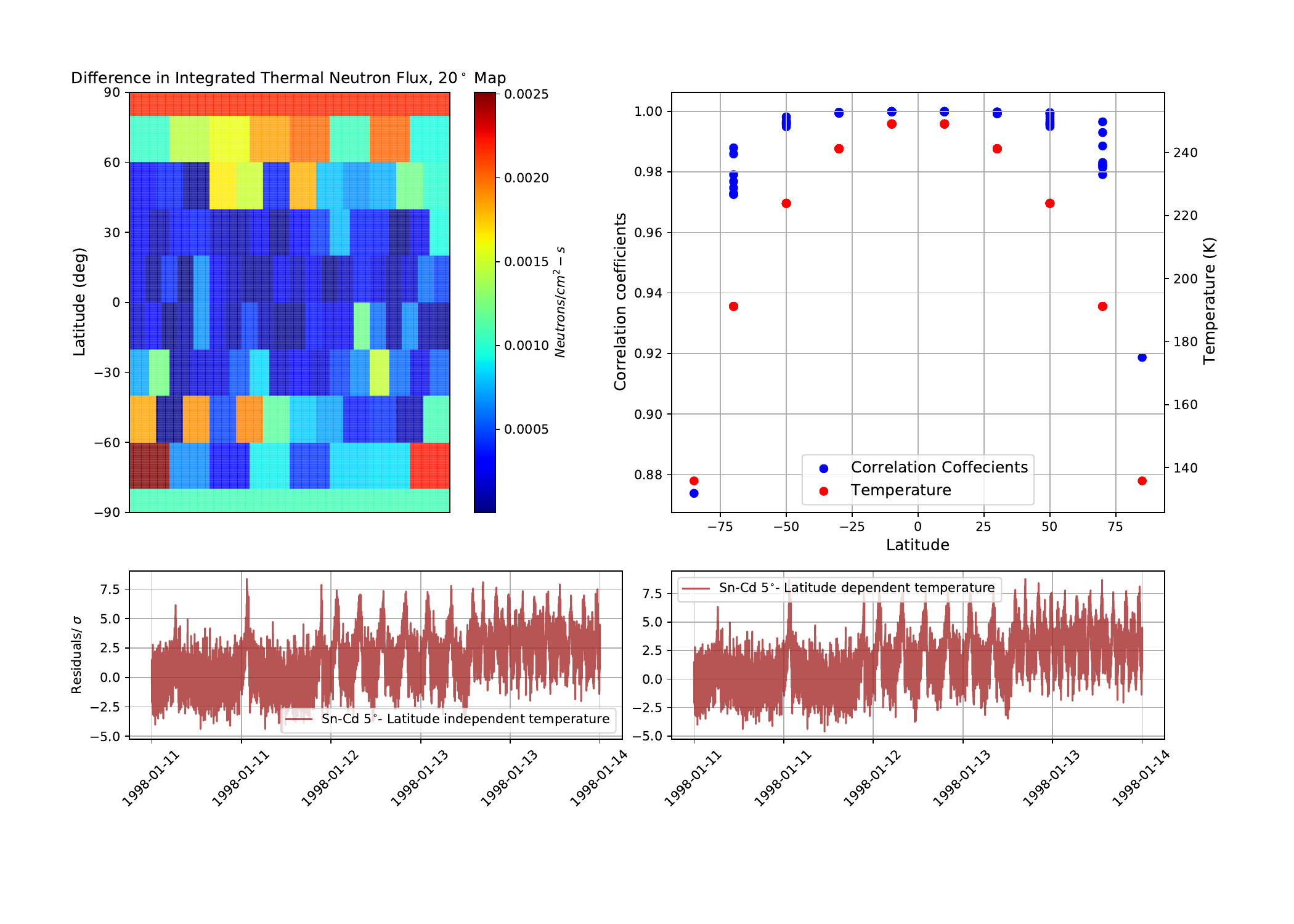}
    \caption{\textit{(Top left:)} Difference in integrated thermal neutrons ($E < 0.4$~MeV) for two different lunar sub-surface temperature conditions - (i) constant equatorial temperature (latitude independent) and (ii) latitude dependent temperature, computed for the 20$^{\circ}$ resolution map. Higher neutron current is seen in at the poles when temperature is set same as that of the equator. \textit{(Top right:)} Correlation coefficients between the two cases. Highest correlation and neutron currents are seen at equators and lowest correlation and neutron currents at the poles where it is colder, indicating the necessity of accounting for the varying surface temperatures in the propagator model. \textit{(Bottom left:)} Residuals in thermal neutrons between LP data and models for latitude independent temperature model and \textit{(Bottom right:)} residuals in thermal neutrons between LP data and models for latitude dependent temperature model. For both the cases, the data at higher altitude sees a lower level of residuals than at lower altitudes.}
    \label{fig:temp_corr}
\end{figure*}

Figure \ref{fig:temp_corr} shows the difference in integrated thermal neutrons, correlation coefficients for neutrons of energy $<0.4~eV$, and the corresponding residuals in thermal neutrons for each of the two temperature conditions. For the calculation of correlation coefficients, a nominal value of $\tau_n$ = 880~s is assumed. Table \ref{tab:temp_effects} gives the RMS in the residuals shown in Figure \ref{fig:temp_corr}. We note that the simple case of constant lunar equatorial temperature results in lower RMS, showing better agreement with LP data compared to latitude dependent temperature model. However, we note that the latitude dependent model is a realistic representation of the lunar subsurface temperatures and such models are therefore necessary in the measurement of neutron lifetime.

\begin{table}[]
    \centering
    \caption{Residuals in thermal neutrons between LP data and models obtained using different lunar subsurface temperatures, evaluated for the 20$^\circ$ map. A lower RMS here indicates better agreement to LP data.}
    \label{tab:temp_effects}
    \begin{ruledtabular}
    \begin{tabular}{c c }
        Temperature condition & RMS in Residuals ($\sigma$) \\
        \hline
        \\
       Constant Equatorial  & \\
       (Latitude independent) & 2.52\\
       \\
       Latitude dependent & 2.68\\
        
    \end{tabular}
    \end{ruledtabular}

\end{table}

\section{Computing Neutron Lifetime}
\label{sec:stats}

We model the neutron propagation to compute the HeCd and HeSn detector count rates, and calculate thermal neutrons. A simple $\chi^2$ statistic is used to find the best-fit neutron lifetime using thermal neutrons as the model parameter. We use $\approx$50 different neutron lifetime values in a broad prior ranging $\tau_n =(540,1680)$~s. This analysis is repeated for each cases of (a) 20$^{\circ}$ resolution, 5$^{\circ}$ resolution, 5 compositions reported in \cite{wilson2021measurement} to study the effect of surface composition; and (b) 20$^{\circ}$ resolution at constant equatorial temperature, and latitude dependent sub-surface temperature model to study the effect of surface temperature on the measurement of neutron lifetime.

\begin{figure*}[]
	\includegraphics[width=\textwidth]{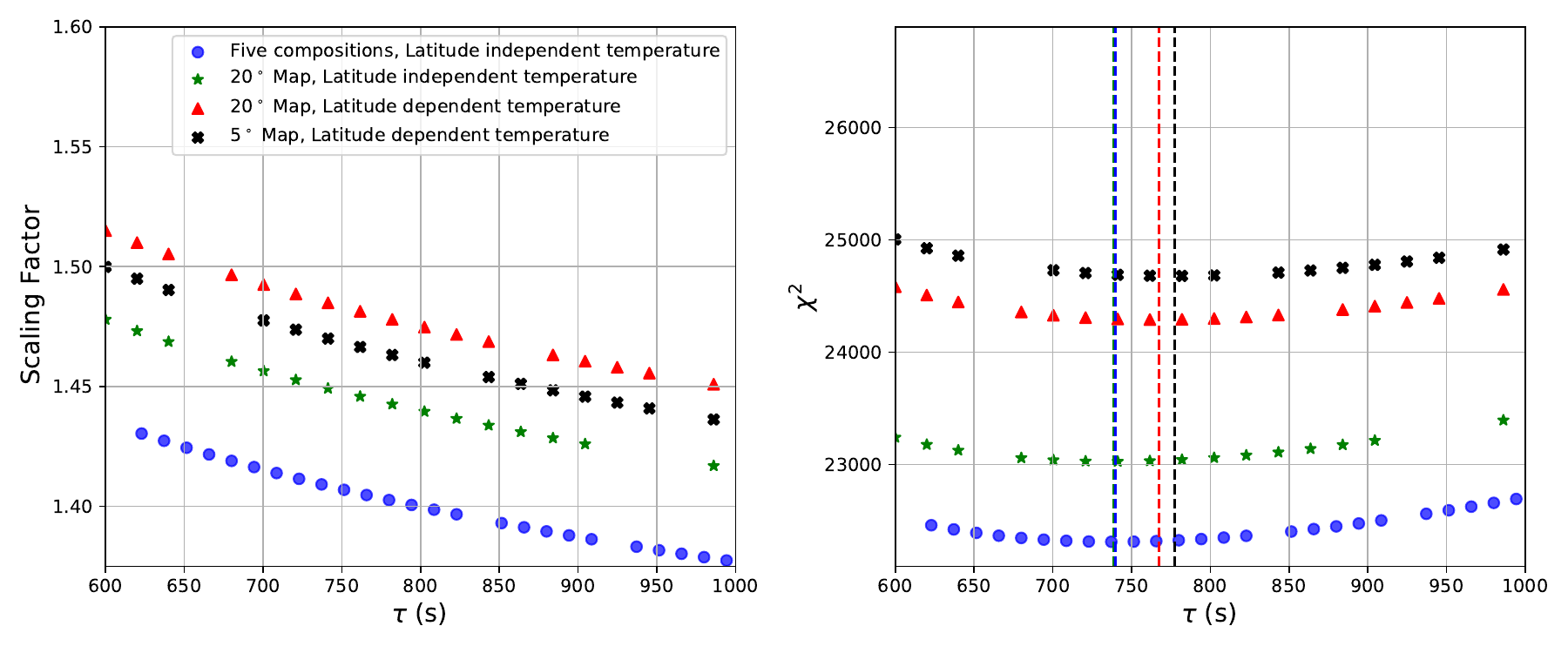}
    \caption{\textit{Left:} Correction factor to account for the model and data offset. Each scaling factor is used as a single multiplicative factor to the thermal neutrons, which is used in $\chi^2$ analysis. The five composition used by \cite{wilson2021measurement} needs the lowest scaling, where as a realistic model capturing complex lunar terrains and latitude dependent temperature effects needs slightly bigger scaling factor. \textit{Right:} $\chi^2$ analysis for the each of the test cases and their corresponding minimum value shown in dashed lines.  }
    \label{fig:scaling_factor}
\end{figure*}

It can be noted from Figure \ref{fig:data_vs_model} that the model is offset from the data, predominantly in the lower altitude regimes. To capture the shortcomings of the model, we scale the thermal neutrons by a constant factor to minimize the discrepancy. The scale factor is constant for a given comparison between model and data, but is recalculated for every $\tau_n$ value. This approach of accounting for the model misfit has also been used by \cite{wilson2021measurement}, however the value of the chosen multiplicative factor has not been reported. Figure \ref{fig:scaling_factor} shows the scaling factor computed for each of the cases capturing lunar composition and temperature effects. The five compositions used by \cite{wilson2021measurement} need the lowest scaling, whereas a model with realistic conditions that capture the complex lunar compositions and temperature effects needs a slightly higher scaling factor. This indicates that a more realistic lunar model of neutron propagation fails to capture the neutron spectra that closely agrees with the LP data. We also note that, with our best understanding of the analysis descriptions and assumptions given in \cite{wilson2021measurement} along with details of the detector response shared through private communications, we fail to retrieve the reported neutron lifetime of $\tau_n = 887\pm 14_{stat}$~s. The notable difference between the two analyses is (i) we use \texttt{MCNP} 6.2,  where as \texttt{MCNPX} has been used in \cite{wilson2021measurement}. (ii) We do not apply temporal offset to the spacecraft altitude and velocity to account for the mismatch in the ephemerides and GRS data. The analysis in \cite{wilson2021measurement} found that the time of the ephemeris did not agree with the time of the count data and required a 900 s offset. We do not find evidence of this offset and suspect that we are using a different version of the ephemerides data. (iii) \cite{wilson2021measurement} analysis is confined to those observations where LP’s longitude was greater than 180$^\circ$ which removed the eastern lunar maria from the study as the residuals in this region were found to be anomalously large. This suggests that the complex composition of this region has not been fully incorporated into their model. These choices lead to an overall lower scaling factor used to offset the model, resulting in better agreement with LPNS data. This work in contrast uses the full orbital insertion data that captures the complex lunar neutron emission fields, and shows the effects of choice of compositional and temperature models on the resulting neutron lifetime measurement.

\begin{table}[]
    \centering
    \caption{Computed $\tau_n$ values along with 1$\sigma$ statistical errors for each of the cases capturing lunar surface composition and temperature effects.}
    \label{tab:tau_values}
    \begin{ruledtabular}
    \begin{tabular}{c c c}
    Map resolution & Temperature & $\tau_n$ (s)\\
    & &\\
    \hline
    \\
        Five compositions  & Latitude independent & 739.6  $\pm$ 10.8
        \\
        re-binned to 2$^\circ$ & & \\
        20$^\circ$ & Latitude independent & 738.6 $ \pm$ 10.8
        \\
       & & \\
        20$^\circ$ & Latitude dependent & 767.3 $\pm$ 11.2
        \\
       & & \\
        5$^\circ$ & Latitude dependent & 777.6 $\pm$ 11.7
        \\
    \end{tabular}
    \end{ruledtabular}
\end{table}

Figure \ref{fig:scaling_factor} also shows the $\chi^2$ analysis for each of the these cases. To determine the neutron lifetime corresponding to least $\chi^2$, we fit a simple parabola and report 1$\sigma$ statistical error-bars on the $\tau_{best-fit}$, which are listed in Table \ref{tab:tau_values}. For a simple five compositions re-binned to 2$^\circ$, $\tau_n = 739.6\pm 10.8$~s, for 20$^\circ$, $\tau_n = 738.6 \pm 10.6$~s when the temperature is latitude independent, and  $\tau_n = 767.3 \pm 11.2$~s assuming latitude dependent temperature. Finally, the 5$^\circ$ map yields $\tau_n = 777.6 \pm 11.7$~s. Using the results from the two temperature models using the 20$^\circ$ map, we calculate the effect of choice of temperature model on $\tau_n$ to be 28.7 $\pm$ 15.5~s. The three maps used here are independent and span over a large parameter space, thus a strict quantification of this choice is inefficacious. However, using the latitude dependent temperature model used in the 20$^\circ$ and 5$^\circ$ maps, we estimate the effect of choice between the two compositional maps to be 10.3 $\pm$ 12.2~s.


By explicitly quantifying the impact of subsurface temperature models and compositional maps on the derived neutron lifetime, we demonstrate that these two systematics alone introduce significant variance, considerably larger than the estimated systematic uncertainties derived in \cite{wilson2021measurement}, yet still fall well short of explaining the full discrepancy between our results and those previously reported. While \cite{wilson2021measurement} applied an empirical scaling factor to reconcile their model with observational data, we find that this approach is insufficient when applied to our independent analysis, further suggesting that additional unaccounted-for systematics are influencing these measurements. These findings highlight a broader issue: that current space-based measurements of the neutron lifetime, particularly those relying on legacy datasets such as from Lunar Prospector, remain subject to large and poorly constrained uncertainties.

\begin{figure}[]
	\includegraphics[width=\columnwidth]{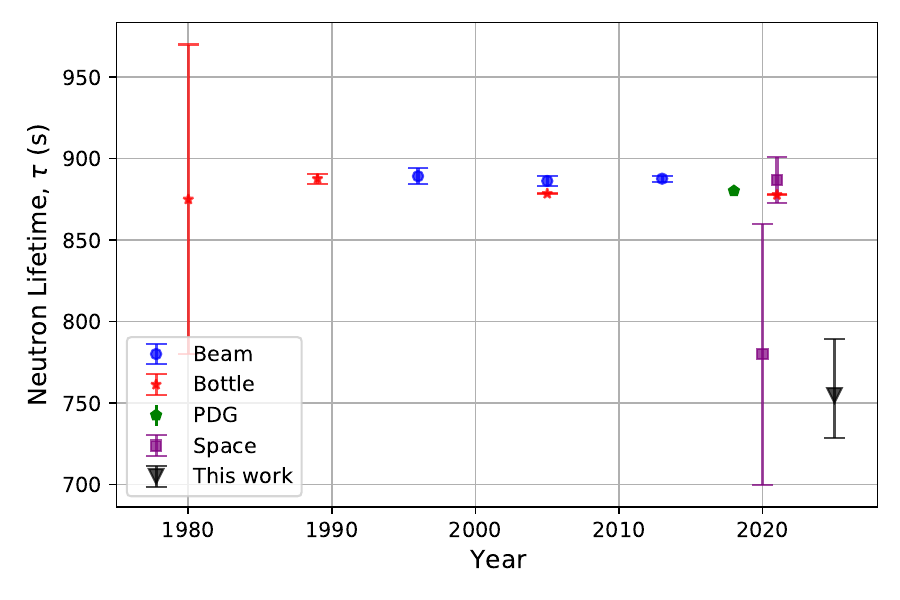}
`    \caption{Reported neutron lifetime values using bottle, beam and space-based techniques in the last five decades. The neutron lifetimes for the four models tested in this work are plotted with their 1$\sigma$ statistical errors. References for PDG - \cite{particle2020review} beam measurements - \cite{byrne1996revised,yue2013improved}, for bottle measurements - \cite{kosvintsev1980use,mampe1989neutron,serebrov2005measurement,gonzalez2021improved} and for space measurements - \cite{wilson2020space, wilson2021measurement}.}
    \label{fig:history}
\end{figure}

We show the results from the four models alongside the reported neutron lifetime for bottle, beam and recent space-based measurements using MESSENGER flybys of Venus and Mercury; and LP data in Figure \ref{fig:history}. Space-based measurements are farther from the current laboratory measurements, however these are intended to serve as a demonstration of the technique using limited non-optimized data. Although surface composition and latitudinal temperature are expected to be significant sources of systematic uncertainties in the measurement, this work is not a comprehensive study of all the systematics. Additional systematics that may not be fully removed by the use of a scaling factor include modeling of the detector response that requires accurate calibration, contributions from uncertainties of simulation physics models, and uncertainties in the GCR spectrum. This work serves to illustrate the effects of choice of the compositional and temperature models on the computed neutron lifetime. A dedicated and rigorous systematic uncertainty analysis, including precise detector calibration and independent lunar surface modeling is essential for future competitive neutron lifetime measurements from space-based neutron data. Such analysis warrants a separate, focused investigation that is beyond the scope of this work.

\section{Future Work}
\label{future}

This work along with recent space-based measurements demonstrate the efficacy of the technique with limited non-optimized data. As these measurements are limited by systematic errors, a competitive measurement using space-based technique demands more data using neutron detectors and orbits specific for neutron lifetime measurements. Currently there are no planned orbital lunar neutron measurements on missions in development or opportunities to propose such a mission. Small missions like LunaH-Map, a cubesat launched in November 2022 on the Space Launch System Artemis-1 mission, could provide an opportunity to demonstrate the collection of neutron data from elliptical lunar orbits and the ability of space-based measurements to quantify the neutron lifetime \cite{hardgrove2020lunar}. Although LunaH-Map was not designed for neutron lifetime measurements, the spacecraft's neutron spectrometer collected lunar neutron and gamma-ray data demonstrating the low-cost pathfinder mission concept for lunar exploration and science \cite{2023LPICo2806.2960H}. Additional work on space-based neutron lifetime measurements will focus on the selection of an appropriate planetary body and orbital parameters for spacecraft insertion among Earth, Moon and Venus, as well as the appropriate instrumentation (i.e. selection of neutron detector). A dedicated mission for measuring the neutron lifetime would have specific orbital requirements for the spacecraft to obtain a larger data-set and achieve a higher confidence measurement of neutron lifetime. The overall increase in data volume, better detector response simulations and modeling of the neutron count rates will help understand and eliminate large systematics, potentially leading to a competitive space-based neutron lifetime measurement. 

The preliminary predicted statistical precision, mission and orbit requirements are laid out in \cite{lawrence2021space}, where Venus is determined to be the best planetary body due to higher statistical precision and simpler surface composition. However, the work lacks study of uncertainties in the neutron lifetime due to its atmospheric composition and temperatures. While Earth is also determined to be a good choice due to ease of orbit insertion and relatively inexpensive mission costs, lunar orbital and landed measurements are equally viable. A detailed study of systematic uncertainties for Venus, Earth and Moon are necessary for a dedicated neutron lifetime mission. This work quantifies the effects of lunar surface composition and temperature but identifies significant unaccounted-for systematics that may explain the large discrepancy from the PDG value. In light of detector calibration limitations, spacecraft geometry effects (e.g., unmodeled shadowing), and the suboptimal nature of the dataset used, future efforts may benefit from the use of improved instrumentation, tailored orbital geometries, more homogeneous targets, and efforts to further evaluate systematic uncertainties that may offer a more robust pathway toward a competitive neutron lifetime measurement with space-based observations.

\section{Conclusion}
\label{sec:conclusion}
We used the HeCd and HeSn detector data during Lunar Prospector's orbit insertion period, where it transitioned from a highly elliptical orbit to a circular one where it was parked for its mission, and computed the neutron lifetime. We used \texttt{MCNP} to get the lunar neutron current, and applied particle kinematics to model the neutron propagation before it hits the LP detector. Thermal neutrons ($E < 0.4$~eV) were used to calculate the neutron lifetime using a $\chi^2$ fit, after a scalar multiplicative correction to account for the misfit between data and model. We also show the effects of precise lunar surface composition and temperature. We find $\tau_n$ to be dependent on surface composition and temperatures, varying from $739.6\pm 10.8$~s, for a simple five compositions re-binned to 2$^\circ$ resolution and assuming a constant equatorial lunar surface temperature, $738.6 \pm 10.8$~s for 20$^\circ$ resolution map and assuming constant temperature, $767.3 \pm 11.2$~s assuming latitude dependent temperature, and $777.6 \pm 11.7$~s for 5$^\circ$ resolution map and assuming latitude dependent temperature. Our analysis underscores the challenges in replicating the neutron lifetime results reported by \cite{wilson2021measurement}, even after extensive efforts to align methodological assumptions and data treatments. The reported measurements here are not yet competitive with the current PDG value pointing to additional systematics that are not yet accounted for. However, this work serves to illustrate the effect on the neutron lifetime due to choice of temperature and compositional models. We estimate the effect on neutron lifetime due to choice of temperature model to be 28.7 $\pm$ 15.5~s, and choice of compositional map (for 20$^\circ$ and 5$^\circ$ maps) to be 10.3 $\pm$ 12.2~s. This analysis was done with a limited non-optimized data, thus showing the potential of such space-based technique for future precise measurements of neutron lifetime.

\section{Acknowledgements}
This research was supported by the LANL through its Center for Space and Earth Science (CSES). CSES is funded by LANL’s Laboratory Directed Research and Development (LDRD) program under project number 20210528CR. This project was conducted as part of the large university collaboration with Arizona State University (ASU). Authors of this paper would like to thank Planetary Science Institute by way of Dr. Bill Feldman and Gavin Nelson for providing the LPNS data; and to Planetary Data System for making the lunar composition maps publicly accessible. Part of this research used HPC and storage resources on Agave super cluster at ASU. This research also used resources provided by the Darwin testbed at LANL which is funded by the Computational Systems and Software Environments subprogram of LANL's Advanced Simulation and Computing program (NNSA/DOE). The authors would like to thank Dr. Brian Weaver and Dr. Arthur Lui for their valuable discussions in developing the preliminary statistical framework. AKV also extends gratitude to Prof. Judd Bowman for his invaluable insights and guidance throughout the analysis. Authors would like to thank the anonymous reviewers of the manuscript.


\newpage

\section{Appendix}
\subsection{Particle Kinematics}

\label{appendic:particle_kinematics}
We take neutrons emitted due to spallation on the lunar surface and add complexities as it travels upwards towards the spacecraft and interacts with the HeCd and HeSn detectors. The piece of the analysis is referred to as \textit{propagation} in this paper. The purpose of propagation is to compute the detector count rates using the neutron current from \texttt{MCNP}, which constitutes as the \textit{model}. The first step in this analysis is to account for the ballistic trajectories of the neutrons due to lunar gravity. For a neutron of energy $K$ and mass $m$, leaving the surface, the binding energy (V) is given by

\begin{equation}
V=\frac{GMm}{R_m}
\end{equation}
where, G is gravitational constant, $R_m$ and $M$ are radius and mass of the Moon respectively. The final energy of the neutron when it hits the detector at a radius of $R$ is given by
\begin{equation}
       K_r = K - V\frac{R-R_m}{R}
\end{equation}
If the neutron leaves the surface at an angle $\theta$ and $\mu = cos\theta$, the incident angle of the particle is given by
\begin{equation}
    \mu_{r}^2 = 1-\Big(\frac{R_m}{R}\Big)^2 \frac{K}{K_r}\big(1-\mu^2\big)
\end{equation}

The second step is to account for the lifetime of the neutron. There could be two cases depending on whether the kinetic energy of the particle is greater or lesser than the binding energy.
When the kinetic energy of the particle is less than the binding energy ($\frac{K}{V}<1$):
\begin{equation}
\begin{split}
\Delta t_r &= \frac {R_M \sqrt{\frac{m}{2V}}} {2(1-\frac{K}{V})^{3/2}} \times \\ 
& \Bigg \{ 2\mu \sqrt{1-\frac{K}{V}} \sqrt{\frac{K}{V}} \Big( 1-\sqrt{\frac{tan^2\theta}{tan^2\theta_R}} \Big)+\\
& sin^{-1}\Big(\frac{B}{\sqrt{A^2 + B^2}}\Big) + sin^{-1} \Big(\frac{1-2K_R/V_R}{\sqrt{A^2 + B^2}}\Big) \Bigg \}
\end{split}
\end{equation}

where $A=\sqrt{4\frac{K}{V}\mu^2(1-\frac{K}{V}) }$ and 
\newline 
$B= 2\frac{K}{V}-1$

When the kinetic energy of the particle is more than the binding energy ($\frac{K}{V}>1$)

\begin{equation}
\begin{split}
    \Delta t_r &= \frac{R_M \sqrt{\frac{m}{2V}}}{2(\frac{K}{V}-1)^{\frac{3}{2}}} \times \\
    & \Bigg\{-2 \mu \sqrt{\frac{K}{V}}\sqrt{\frac{K}{V}-1}\Big(1-\sqrt{\frac{\textrm{tan}^2 \theta}{\textrm{tan}^2 \theta_R}}\Big)\\
    & -\textrm{ln}\Big[\frac{ 2 \mu \sqrt{\frac{K}{V}}\sqrt{\frac{K}{V} - 1}  \sqrt{\frac{\textrm{tan}^2 \theta}{\textrm{tan}^2 \theta_R}} + (2\frac{K_r}{V_r}-1)}{ 2 \mu \sqrt{\frac{K}{V}} \sqrt{\frac{K}{V} - 1 } + (2\frac{K}{V}-1)}\Big] \Bigg\}
\end{split}
\end{equation}

The third step is to apply effective area of the detector and account for the geometric factor correction caused due to solid angle effects at the altitude of the spacecraft,

\begin{equation}
    \Omega(h) = 2\pi \bigg\{1 - \sqrt{1-\frac{R^2}{(R+h)^2}}\bigg\}
\end{equation}

\subsection{True Anomaly}
\label{app:true_anomaly}

The location of neutron emission is calculated by treating the neutron motion as orbital motion. We calculate the angle between the neutron periapsis and current location, that is, the true anomaly, at both the satellite position and the lunar surface. The difference between the true anomalies at these locations is the difference in central angle between neutron emission and the satellite location. For example, for a neutron travelling due north, the difference in true anomaly simply gives the difference in latitude between the satellite and emission location. For other neutron directions, this angle is rotated.

We calculate the true anomaly following the notation of \cite{Curtis_OMES2005}. Note that some symbols are reused from the previous section with different meanings. Defining the gravitational parameter $\mu = G M$ for the gravitational constant $G$ and mass of the Moon $M$ and the magnitude of the specific relative angular momentum $h=rv_{\perp}$ for radius $r$ and perpendicular component of the velocity $v_{\perp}$, the eccentricity is:

\begin{equation}
    e=\frac{1}{\mu}\sqrt{(2\mu-rv^{2})rv_{r}^{2}+(\mu-rv^{2})^2}.
\end{equation}

$v_{r}$ is the radial component of the velocity. $e$ is a constant and can be evaluated at any point in the orbit; in practice we evaluate it at the point of emission. Then the true anomaly at any radius $r$ is:

\begin{equation}
    \theta=\textrm{arccos}\Big[\frac{1}{e}(\frac{h^{2}}{\mu r}-1)\Big].
\end{equation}

For gravitationally-bound neutrons on inward trajectories when they intersect the satellite, the true anomaly is $2 \pi - \theta$ as calculated in the equation above.

\subsection{Co-ordinate Transformation}
\label{appendix:transforms}
Following the kinematics described in Section \ref{comp_prop_model}, we get the neutron energy in a frame of reference where $\hat{z_r}$ is towards the local zenith at the spacecraft. To get the thermal velocity of the neutron in the rest frame of the spacecraft, we should transform the coordinate system where $\hat{z}$ is aligned with the rotation axis of the Moon and the spacecraft.

Assuming $\hat{y_r}$ to lie entirely in the x-y plane in the standard Moon frame, we can define the spacecraft position by $(x_s, y_s, z_s)$, such that radius of the spacecraft is given by $R=\sqrt{x^2_s + y^2_s + z^2_s}$ and the component in the x-y plane is $r=\sqrt{x^2_s + y^2_s}$.

The velocity of the spacecraft is thus defined by:

\begin{equation}
\vec{v} = v_{x,r}\hat{x_r} + v_{y,r}\hat{y_r} + v_{z,r}\hat{z_r}
\end{equation}

\begin{equation}
\begin{split}
\vec{v}   &= \bigg[ v_{x,r} \frac {x_s z_s}{rR} + v_{y,r}\frac{-y_s}{r} + v_{z,r}\frac{x_s}{R}\bigg] \hat{x} \\
    &\quad + \bigg[ v_{x,r} \frac {y_s z_s}{rR} + v_{y,r}\frac{x_s}{r}  + v_{z,r}\frac{y_s}{R}\bigg] \hat{y} \\
    &+ \bigg[ v_{x,r} \frac {-r^2}{rR} + v_{z,r}\frac{z_s}{R}\bigg] \hat{z}
\end{split}
\end{equation}

We then subtract spacecraft velocity to get neutron velocity in the spacecraft rest and rotation frame, and recalculate energy and angle.

\newpage
\nocite{*}
\bibliography{references}{}

\end{document}